\documentclass{ws-p8-50x6-00}

\newcommand{\ba}{\begin{eqnarray}}
\newcommand{\ea}{\end{eqnarray}}

\begin{document}

\title{$K_{\ell4}$ at two-loops and CHPT predictions for $\pi\pi$-scattering}

\author{Gabriel Amor\'os, Johan Bijnens}

\address{Department of Theoretical Physics 2, Lund University\\
S\"olvegatan 14A, S 22362 Lund, Sweden}

\author{Pere Talavera}

\address{Groupe de Physique Th\'eorique,
Institut de Physique Nucl\'eaire,\\
Universit\'e Paris Sud,
F-91406 Orsay Cedex, France}

\maketitle

\abstracts{We present the results from our two-loop calculations
of masses, decay-constants, vacuum-expectation-values
and the $K_{\ell4}$ form-factors in three-flavour Chiral Perturbation
Theory (CHPT).
We use this to fit the $L_i^r$ to two-loops and discuss the ensuing
predictions for $\pi\pi$-threshold parameters.} 

\enlargethispage{1cm}
\footnotetext{$^*$Talk presented at Chiral Dynamics 2000, Jefferson Lab,
Newport News,
Virginia, USA, July 17-22, 2000.}

\vskip-10cm \phantom{p}\hfill LU TP 00-46\\ \phantom{p}\hfill hep-ph/0011023\\
\phantom{p}\hfill October 2000
\vskip8cm

We present the main results of a rather long program of
two-loop calculations in three flavour CHPT in the meson sector.
The Lagrangian at lowest order $p^2$ has two-parameters and two quark masses.
In practice, in the isospin limit, this leads to $F$, $B_0 \hat{m}$
and $B_0 {m_s}$ as the three free parameters. At order $p^4$
there are an additional\cite{GL1} 10 and at order $p^6$ there are\cite{BCElag}
90. An overview of the calculations at two-loops in
the two-flavour sector can be found in \cite{Mainz}, presented in the
previous Chiral Dynamics meeting. Since then the 
$\pi\pi$ scattering\cite{pipi},
the vector and scalar form-factor\cite{BCT}
and the expressions in terms of the $p^6$ low-energy
constants have been published\cite{BCEinf}. In the three-flavour
sector the generalized double-log contributions were found to
be large\cite{BCEdbl}. This prompted us to start our
program so a major refit of CHPT parameters to two-loop order would be
possible. Other three flavour $p^6$ work is in Ref.\cite{Other}.


In Ref.\cite{ABT1} we calculated the $VV$ and $AA$ two-point functions
to two-loop order and in Refs.\cite{ABT2,ABT3} the $K_{\ell4}$
form-factors and the vacuum-expectation-values. Future work is concentrating
on isospin violating effects.

\begin{figure}
\footnotesize
\begin{minipage}{0.61\textwidth}
\caption{The fit of $L_i^r$ described in the text.
We have shown some of the input variations used.
\label{tabLi}}
\begin{tabular}{|ccccc|}
\hline
		& Main Fit &fit 2&
  fit 8 & fit 9\\
\hline
$10^3\,L^r_1$&   0.52$\pm$0.23&   0.53&
   0.63&   0.65\\	       
$10^3\,L^r_2$&   0.72$\pm$0.24&   0.73&
   0.73&   0.85\\      
$10^3\,L^r_3$&$-$2.70$\pm$0.99&$-$2.71&
$-$2.67&$-$3.27\\	       
$10^3\,L^r_5$&   0.65$\pm$0.12&   0.62&
   0.51&   0.60\\ 	       
$10^3\,L^r_7$&$-$0.26$\pm$0.15&$-$0.20&
$-$0.25&$-$0.26\\	       
$10^3\,L^r_8$&   0.47$\pm$0.18&   0.35&
   0.44&   0.48\\
\hline	     
changed      & &$m_s/\hat m$&
$\mu$&$g(0)$\\ 
quantity     &          & 26  
&1.0&4.93\\
\hline
\end{tabular}
\end{minipage}
\begin{minipage}{0.39\textwidth}
\caption{The results for $\pi\pi$ scattering and
 the new data  (Pislak).\label{figpipi}}
\epsfxsize=\textwidth
\epsfbox{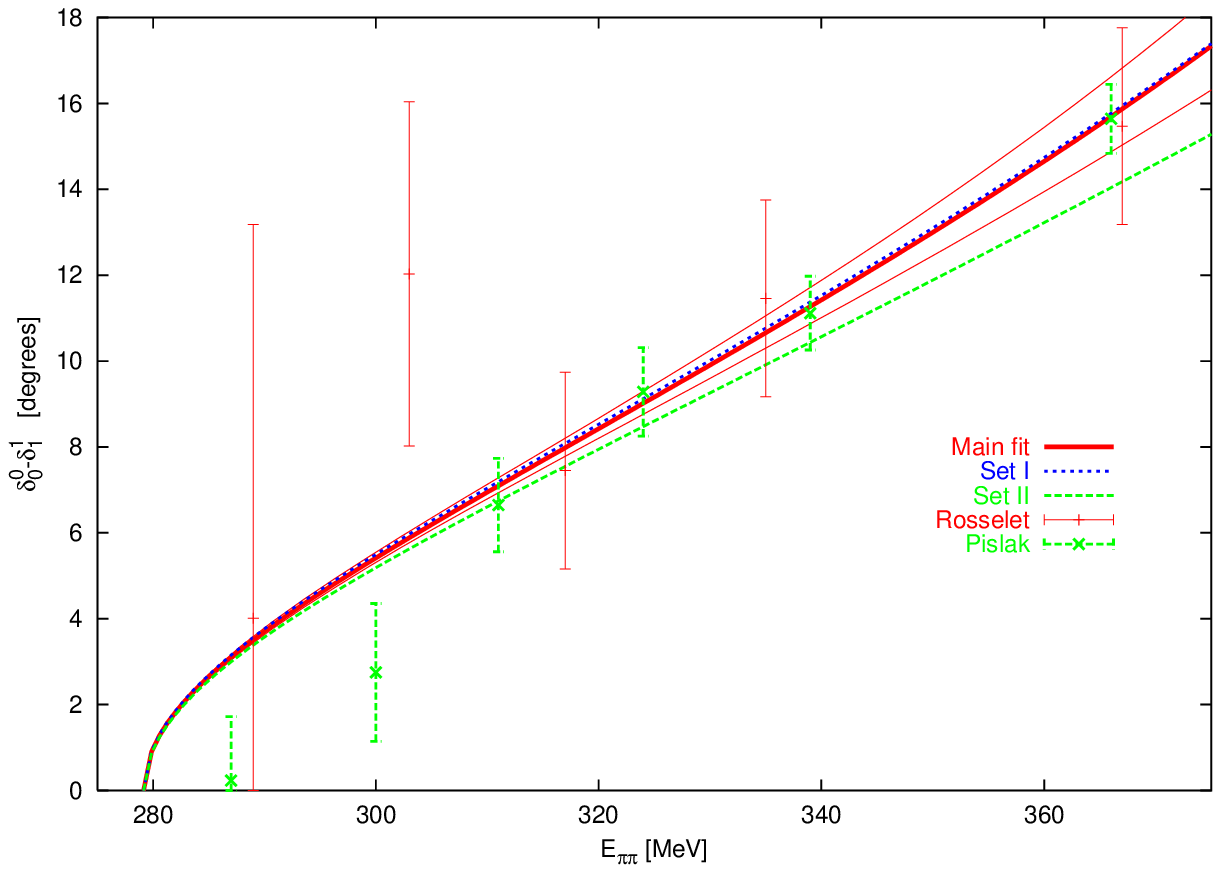} 
\end{minipage}
\end{figure}

There is a large number of parameters
that needs determination. We have been forced to use assumptions to
estimate the $p^6$ constants but checked that varying them within a factor
of two didn't change results significantly. In addition we
used $L_4^r=L_6^r=0$ and $m_s/\hat{m}=24$ as theoretical input.
Experimental input is $F_\pi$, $F_K/F_\pi$, $M_\pi$, $M_K$,
$M_\eta$ and from $K_{\ell4}$, $F(0)$, $G(0)$ and the slope $\lambda$.
The resulting fit for the $L_i^r$ is quite stable. 
Fit results without the $\eta^\prime$ contribution in the $p^6$ estimate
are in \cite{ABT2} and with it in \cite{ABT3}.
The main results and the variation with some of the input
are shown in Fig. \ref{tabLi}. Notice the variation with $K_{\ell4}$ input
in fit 9. 
We look forward to more $K_{\ell4}$ data that would add more
precision, especially in the value of the slopes for $F$ and $G$, to have more
redundancy in the input.
The excellent
prediction for $\pi\pi$ is shown with data,
including the new BLN data\cite{Zeller}, in Fig.~\ref{figpipi}.

\section*{\vskip-0.7cm Acknowledgments\vskip-0.2cm}

Work partially supported by the EU TMR Network
EURODAPHNE (Contract No. ERBFMX-CT98-0169) and by the Swedish Science
Foundation.

\end{document}